\documentclass[journal]{IEEEtran}
\usepackage{amsmath,graphicx}
\usepackage{amsfonts,amssymb}
\usepackage{multirow}
\usepackage{algorithm}
\usepackage{algpseudocode}
\usepackage{cite}
\usepackage{subfig}
\DeclareGraphicsExtensions{.jpg,pdf,.png}

\ifCLASSINFOpdf
  
\else

\fi
\begin{document}
\title{Optimum Decoder for Multiplicative Spread Spectrum Image Watermarking with Laplacian Modeling}
\author{Nematollah Zarmehi${}^{1,2,*}$\thanks{* Corresponding author. \newline Email addresses: {zarmehi\_n@ee.sharif.edu} (N. Zarmehi), {aref@sharif.edu} (M.R. Aref)} and Mohammad Reza Aref${}^{1,3}$\\
	{\small{${}^1$Electrical Eng. Department, Sharif University of Technology, Tehran, Iran\\ ${}^2$Advanced Communications Research Institute, Sharif University of Technology, Tehran, Iran\\ ${}^3$Information Systems and Security Lab. (ISSL), Electrical Eng. Department, Sharif University of Technology, Tehran, Iran}}}
\maketitle
\begin{abstract}
This paper investigates the multiplicative spread spectrum watermarking method for the image. The information bit is spreaded into middle-frequency Discrete Cosine Transform (DCT) coefficients of each block of an image using a generated pseudo-random sequence. Unlike the conventional signal modeling, we suppose that both signal and noise are distributed with Laplacian distribution because the sample loss of digital media can be better modeled with this distribution than the Gaussian one. We derive the optimum decoder for the proposed embedding method thanks to the maximum likelihood decoding scheme. We also analyze our watermarking system in the presence of noise and provide analytical evaluations and several simulations. The results show that it has the suitable performance and transparency required for watermarking applications.
\end{abstract}

\IEEEpeerreviewmaketitle
\section{Introduction}\label{sec:intro}
\IEEEPARstart{T}{oday}, copyright infringement, unauthorized publications, and the proof of ownership are the main concerns of digital media producers. These concerns in the digital media business are growing as the technology advances. It is not difficult to illegally publish a large amount of digital media on the Internet in few seconds. Watermarking is a solution for these issues \cite{ref:book-cox}.

Watermarking is a branch of data hiding in which a watermark is added to the media for different purposes. For example, the watermark can protect media from illegal copying. It can also be used for detecting the unauthorized publications and proof of ownership. It can be applied to digital media such as text, audio, image, and video \cite{ref:akhaee-invited-survey}.

Spread spectrum embedding scheme is one of the most famous embedding methods proposed by Cox {\it{et al.}} in 1997 \cite{ref:cox-spread}. In this method, the embedder spreads the information bits into a cover media using a pseudo-random sequence. On the other side, the extractor despreads the information bits from that watermarked media. By this approach, the watermarked signal has a high level of robustness against some attacks. The spread spectrum methods are divided into two major categories: additive and multiplicative. As the names of these two categories indicate, in additive methods, we have the summation of the cover signal with a pseudo-random sequence while in multiplicative methods, we have the multiplication of them. Multiplicative methods have higher robustness than the additive ones \cite{ref:zarmehi-eusipco}-\cite{nematiet}. Several additive \cite{ref:addnikolas,ref:addzaidi,ref:addhuang,hernandez,Stouraitis} and multiplicative \cite{ref:mjidong,ref:mcampisi,ref:mbarni,ref:mkirovski,ref:mvalizadeh} watermarking methods are proposed in this area of research.

In \cite{ref:siglaplace}, a multiplicative watermarking method is proposed for audio and speech signals in which the host signal and noise are modeled with Laplacian and Gaussian distributions, respectively. A DCT-based multiplicative watermarking method is also proposed in \cite{cauchy} where the host signal is modeled by Cauchy distribution and the optimal detector has been derived in the absence of noise.

The impact of secure watermark embedding in digital images studied in \cite{transport}. Authors proposed a practical implementation of secure spread spectrum watermarking using distortion optimization. The host signal in this work is modeled by Gaussian distribution.

In \cite{ref:zarmehi-iscisc}, an additive spread spectrum method is investigated. In this work, both signal and noise are modeled with Laplacian distributions. Moreover, the embedding rule is so simple.

In this paper, we investigate the conventional multiplicative spread spectrum watermarking in transform domain. An optimum decoder is also derived using the Maximum Likelihood Decoding (MLD) technique. We suppose that both signal and noise are distributed with Laplacian distributions. In digital media transmission, we are facing with sample loss or corruption. In these applications, the Gaussian distribution cannot be a good model for noise. The sample loss and corruption are usually modeled with impulsive noise \cite{ref:farsiu}, which needs a larger heavy-tail distribution than Gaussian one \cite{ref:markslaplace}. We will provide comparisons between our scheme and the conventional scheme where the signal and noise are modeled with Gaussian distributions.

This paper is organized as follows: In Section \ref{sec:sig-model}, we discuss suitable statistical modeling for signal and noise. Section \ref{sec:wat} is devoted to our watermarking system where the embedder and optimum decoder are presented. We analyze our watermarking system in presence of noise in Section \ref{sec:noise}. Some analytical evaluations and simulation results are presented in Section \ref{sec:sim} and finally Section \ref{sec:conclusion} concludes the paper.

\section{Signal Modeling}\label{sec:sig-model}
We embed the information bits into Discrete Cosine Transform (DCT) coefficients of the cover media. The Generalized Gaussian Distribution (GGD) with small shape parameter is a suitable statistical modeling for DCT coefficients. The GGD is stated as follows:
\begin{equation}
f_X(x)=Ae^{-|\beta (x-m)|^c}
\end{equation}
where
\begin{equation}
\beta = \frac{1}{\sigma}\sqrt{\frac{\Gamma(3/c)}{\Gamma(1/c)}}, \quad A = \frac{\beta c}{2\Gamma(1/c)},
\end{equation}
$m$ is the mean, $\sigma$ is the standard deviation, $c$ is the shape parameter, and $\Gamma(.)$ is the Gamma function.

A suitable and yet simple special case of GGD is the Laplacian distribution which is GGD with $c=1$ \cite{ref:transfcoding}. We model the DCT coefficients with a Laplacian distribution. In some applications, the noise is modeled as a Gaussian random variable. But in some cases, for example corrupted pixels of an image are usually modeled with impulsive noise \cite{ref:farsiu}. The impulsive noise needs a distribution function with larger heavy-tail than the Gaussian distribution \cite{ref:markslaplace}. Therefore, we also consider Laplacian distribution for noise in this paper. We will verify our claim in Subsection \ref{subsec:sigdist}.

\section{Proposed Method}\label{sec:wat}
In this section, the embedding method and its optimum decoder are proposed.

\subsection{Watermark Embedding}\label{subsec:embedder}
\begin{figure}[t!]
	\centering
	\includegraphics[width=0.45\linewidth]{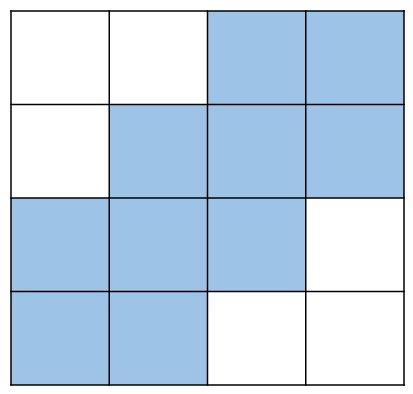}
	\caption{The middle frequency coefficients of $4\times 4$ block that are selected for embedding.}
	\label{fig:middle}
\end{figure}

Our goal is to embed each information bit $b\in\{0,1\}$ into $N$ middle frequency DCT coefficients of each $4\times 4$ blocks of the cover signal. The selected middle frequency coefficients are showed in Fig. \ref{fig:middle}. First, a pseudo-random sequence $\mathbf{w} = \left[{w_1,w_2,...,w_N}\right]^T$ is generated and the information bit $b$ is embedded as follows:
\begin{equation}\label{eq:embd}
\mathbf{y} = \mathbf{x}.* \left[{1+\alpha(2b-1)\mathbf{w}}\right]
\end{equation}
where $\mathbf{x} = \left[{x_1,x_2,...,x_N}\right]^T$ is the cover signal, $\alpha$ is the strength factor, and $\mathbf{y} = \left[{y_1,y_2,...,y_N}\right]^T$ is the watermarked sequence. Here, it is assumed that the elements of the pseudo-random sequence is selected from a binary set $w_i\in\{-1,+1\}$ with equal probability. The symbol $.*$ is used for vector multiplication. For example, $\mathbf{u}.*\mathbf{v}$ denotes the element-by-element product of $\mathbf{u}$ and $\mathbf{v}$. According to (\ref{eq:embd}), the watermark power can be controlled by the strength factor. The more value of strength factor, the more power of the watermark signal and therefore the more distortion.

\subsection{Watermark Decoding}\label{subsec:decoder}
Consider the following binary statistical hypothesis test for extracting the watermark bit.
\begin{equation}
\text{Hypothesis Test 1} = \left\{ {\begin{array}{*{20}{c}}
	{{H_0}:}&{b = 0}\\
	{{H_1}:}&{b = 1}
	\end{array}} \right.
\end{equation}
The cover signal is assumed to be distributed as i.i.d. Laplacian distribution with parameter $\lambda_x$ as follows:
\begin{equation}
f_X(x_i)=\frac{\lambda_x}{2}e^{-\lambda_x|x_i|}, \quad i=1,2,...,N.
\end{equation}
According to (\ref{eq:embd}), the distribution of the watermarked signal conditioned to the information bit, $b$, can be computed as
\begin{equation}\label{eq:fycond}
f_{Y|b}(y_i)= \frac{\lambda_xe^{-\lambda_x|\frac{y_i}{1+\alpha(2b-1)w_i}|}}{2(1+\alpha(2b-1)w_i)} \quad i=1,2,...,N.
\end{equation}

\begin{figure*}[t!]
	\begin{minipage}[t]{1\linewidth}
		\centering
		\centerline{\includegraphics[width=0.7\linewidth]{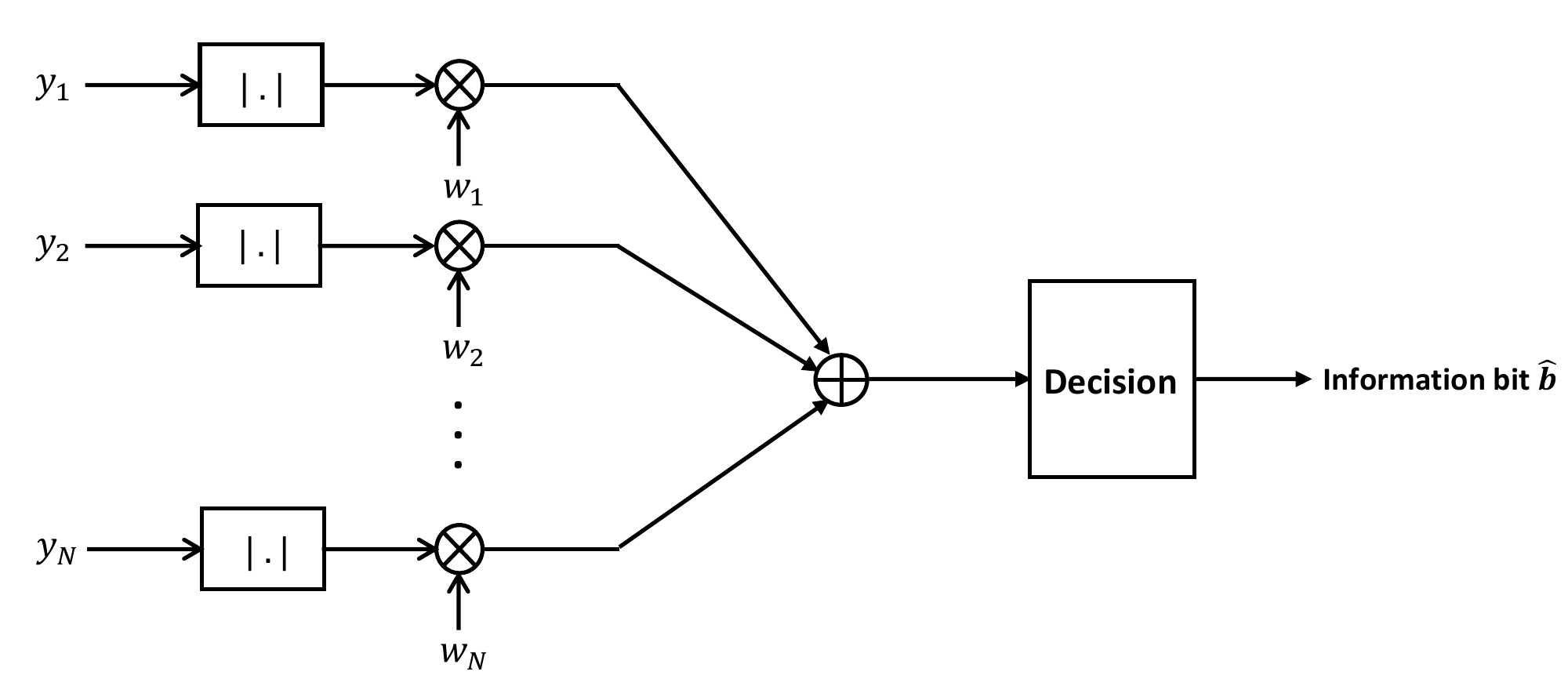}}
		\centerline{ }\medskip
	\end{minipage}
	\caption{Block diagram of the optimum decoder in noise free environment.}
	\label{fig:blck1}
\end{figure*}

We consider that same probabilities for bits 0 and 1, Therefore, MLD is optimal. In order to extract the watermark bit, we employ the optimum decoder based on the MLD scheme
\begin{equation}
\hat{b} = \mathop {\arg \max }\limits_{ b\in \{0,1\} }{f(\mathbf{y}|b)}.
\end{equation}
We make the likelihood function as follows
\begin{equation}\label{eq:ml10}
L(\mathbf{y}) =  \frac{f(\mathbf{y}|b=1)}{f(\mathbf{y}|b=0)} = \frac{\prod\limits_{i=1}^{N}{f_{Y|b=1}(y_i)}}{\prod\limits_{i=1}^{N}{f_{Y|b=0}(y_i)}}
\begin{array}{*{20}{c}}
{H_1}\\
{\gtrless}\\
{H_0}
\end{array}
1.
\end{equation}
Substituting (\ref{eq:fycond}) in (\ref{eq:ml10}), we have:
\begin{align}\label{eq:ml1}
\begin{split}
L(\mathbf{y}) = {}& \frac{\prod\limits_{i=1}^{N}{f_{Y|b=1}(y_i)}}{\prod\limits_{i=1}^{N}{f_{Y|b=0}(y_i)}} \\
={}& \prod\limits_{i=1}^{N}{\frac{\frac{1}{1+\alpha w_i}e^{-\lambda_x|\frac{y_i}{1+\alpha w_i}|}}{\frac{1}{1-\alpha w_i}e^{-\lambda_x|\frac{y_i}{1-\alpha w_i}|}}} 
\begin{array}{*{20}{c}}
{H_1}\\
{\gtrless}\\
{H_0}
\end{array}
1.
\end{split}
\end{align}
Since $L(\mathbf{Y})$ is positive and logarithmic function is strictly increasing we can take the logarithm of (\ref{eq:ml1}) as follows:

\begin{align}
\begin{split}
\sum\limits_{i=1}^{N}{\left\{{\log\left({\frac{1-\alpha w_i}{1+\alpha w_i}}\right) + \lambda_x \left[{\left|{\frac{y_i}{1-\alpha w_i}}\right| - \left|{\frac{y_i}{1+\alpha w_i}}\right|}\right]}\right\}} & ~ \\
\begin{array}{*{20}{c}}
{H_1}\\
{\gtrless}\\
{H_0}
\end{array} & 
0.
\end{split}
\end{align}
After some mathematical manipulations, the optimum decision rule is simplified as
\begin{equation}
T
\begin{array}{*{20}{c}}
{H_1}\\
{\gtrless}\\
{H_0}
\end{array}  
\tau,
\end{equation}
where
\begin{align}
T = \sum\limits_{i=1}^{N}{|y_i|w_i} 
\quad \mbox{and} \quad
\tau= \frac{1-\alpha^2}{2\alpha\lambda_x}\sum\limits_{i=1}^{N}{ w_i\log\left({\frac{1+\alpha}{1-\alpha}}\right)  }.
\end{align}

Block diagram of the optimum decoder is shown in Fig. \ref{fig:blck1}. In decoder side, the absolute sum of the received samples are multiplied by the pseudo-random sequence $\mathbf{w}$ and the result is compared with a threshold, $\tau$.

\section{Noise Analysis}\label{sec:noise}
In this section, we analyze our watermarking system in the presence of noise. The decoder receives the watermarked signal plus channel noise
\begin{equation}
\mathbf{z} = \mathbf{y} + \mathbf{n}
\end{equation}
Here, we suppose that the channel noise $\mathbf{n}=[n_1,n_2,...,n_N]^T$ are additive i.i.d. Laplacian random variables with parameter $\lambda_n$. Moreover, we suppose that channel noise is independent of the cover signal. Hence, the distribution of the received signal, $\mathbf{z}$, is convolution of two Laplacian distributions with parameters $\lambda_y$ and $\lambda_n$ and can be obtained as \cite{ref:zarmehi-iscisc}
\begin{equation}\label{eq:fz}
f_Z(z_i) = \frac{1}{2}\frac{\lambda_y}{\lambda_n}\frac{1}{1-\left({\frac{\lambda_y}{\lambda_n}}\right)^2}\left[{
	\frac{1}{\lambda_y}e^{-\frac{|z_i|}{\lambda_n}} - \frac{1}{\lambda_n}e^{-\frac{|z_i|}{\lambda_y}}
	}\right].
\end{equation}
As previous section, we again make the following binary hypothesis test for extracting the information bit
\begin{equation}
\text{Hypothesis Test 2} = \left\{ {\begin{array}{*{20}{c}}
	{{H_0}:}&{b = 0}\\
	{{H_1}:}&{b = 1}
	\end{array}} \right.
\end{equation}

\begin{figure*}[t!]
	\begin{minipage}[b]{1\linewidth}
		\centering
		\centerline{\includegraphics[width=0.7\linewidth]{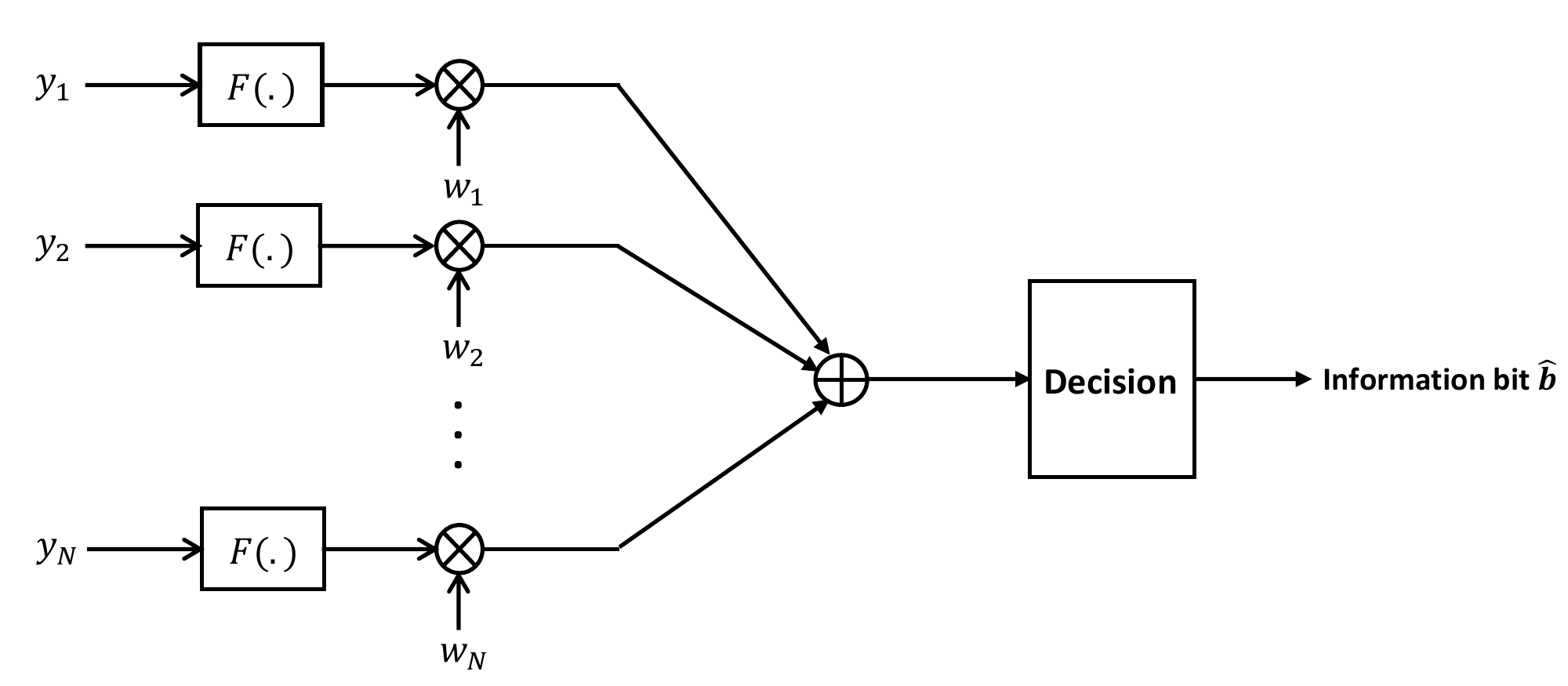}}
		\centerline{ }\medskip
	\end{minipage}
	\caption{Block diagram of the optimum decoder in presence of noise.}
	\label{fig:blck2}
\end{figure*}

Again by employing the MLD scheme for above test we have
\begin{equation*}
\hat{b} = \mathop {\arg \max }\limits_{ b\in \{0,1\} }{f(\mathbf{z}|b)},
\end{equation*}
\begin{equation}\label{eq:ml2}
L(\mathbf{z}) = {} \frac{f(\mathbf{z}|b=1)}{f(\mathbf{z}|b=0)} = \frac{\prod\limits_{i=1}^{N}{f(z_i|b=1)}}{\prod\limits_{i=1}^{N}{f(z_i|b=0)}}
\begin{array}{*{20}{c}}
{H_1}\\
{\gtrless}\\
{H_0}
\end{array}
1.
\end{equation}
Substituting (\ref{eq:fz}) in (\ref{eq:ml2}), we get
\begin{align}\label{eq:ml23}
\begin{split}
\prod\limits_{i=1}^{N} &
\left\{ \frac{\frac{\lambda_x}{\lambda_n(1+\alpha w_i)}\frac{1}{1-\left({\frac{\lambda_x}{\lambda_n(1+\alpha w_n)}}\right)^2}}{\frac{\lambda_x}{\lambda_n(1-\alpha w_i)}\frac{1}{1-\left({\frac{\lambda_x}{\lambda_n(1-\alpha w_i)}}\right)^2}} \times \right.\\
 ~&\left. \frac{		\frac{1+\alpha w_i}{\lambda_x}e^{-\frac{|z_i|}{\lambda_n}} - \frac{1}{\lambda_n}e^{-\frac{|z_i|(1+\alpha w_i)}{\lambda_x}}
	}{	\frac{1-\alpha w_i}{\lambda_x}e^{-\frac{|z_i|}{\lambda_n}} - \frac{1}{\lambda_n}e^{-\frac{|z_i|(1-\alpha w_i)}{\lambda_x}}}
	\vphantom{\frac{\frac{\lambda_x}{\lambda_n(1+\alpha w_i)}\frac{1}{1-\left({\frac{\lambda_x}{\lambda_n(1+\alpha w_n)}}\right)^2}}{\frac{\lambda_x}{\lambda_n(1-\alpha w_i)}\frac{1}{1-\left({\frac{\lambda_x}{\lambda_n(1-\alpha w_i)}}\right)^2}}}
	\right\}
\begin{array}{*{20}{c}}
{H_1}\\
{\gtrless}\\
{H_0}
\end{array}
1.
\end{split}
\end{align}
Again taking logarithmic function from both sides of (\ref{eq:ml23}) and doing some simplifications, it can be rewritten as
\begin{equation}
\sum\limits_{i=1}^{N}{w_iF(z_i,\lambda_x,\lambda_n)} \begin{array}{*{20}{c}}
{H_1}\\
{\gtrless}\\
{H_0}
\end{array}
\tau_n
\end{equation}
where
\begin{equation}
F(z_i,\lambda_x,\lambda_n) =  \log\left({\frac{1-\frac{\lambda_x}{\lambda_n(1+\alpha)}e^{|z_i|\left({\frac{1}{\lambda_n}-\frac{1+\alpha}{\lambda_x}}\right)}}{1-\frac{\lambda_x}{\lambda_n(1-\alpha)}e^{|z_i|\left({\frac{1}{\lambda_n}-\frac{1-\alpha}{\lambda_x}}\right)}}}\right),
\end{equation}
and
\begin{equation}
\tau_n = \sum\limits_{i=1}^{N}{w_i\log\left[{\left({\frac{1-\alpha}{1+\alpha}}\right)^2\left({\frac{(1+\alpha)-(\lambda_x/\lambda_n)^2}{(1-\alpha)-(\lambda_x/\lambda_n)^2}}\right)}\right]}.
\end{equation}

Fig. \ref{fig:blck2} shows the block diagram of the optimum decoder in presence of noise. It is obvious that this decoder is more complex than the decoder in noise free environment. But we will see that we get much more gain in probability of error than the case the signal and noise are modeled with Gaussian distributions.

\section{Simulation Results}\label{sec:sim}
In this section, we present some analytical evaluations and the simulation results. First, we show that the Laplacian distribution is a proper approximation for the DCT coefficients of the image. Then, the transparency of the proposed watermarking method is evaluated. Finally, we evaluate the performance of our watermarking method and its robustness against some attacks and compare it with the case in which both signal and noise are modeled with Gaussian distributions.

\subsection{Signal Distribution}\label{subsec:sigdist}
\begin{figure}[h!]
	\centering
	\subfloat[~\label{subfig:fit1}]{%
		\includegraphics[width=0.25\textwidth]{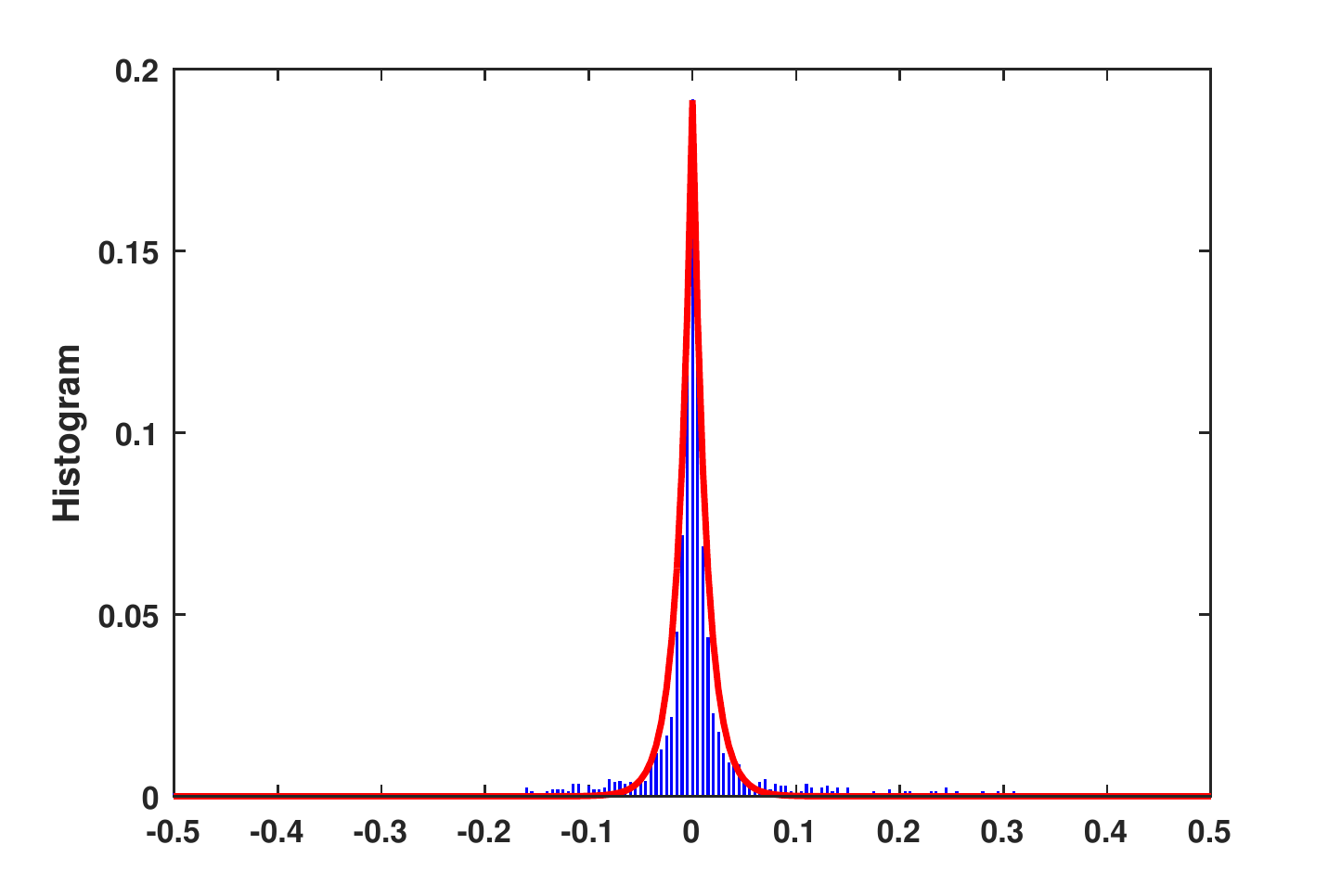}
	}
	\hspace{-0.58 cm}
	\subfloat[~\label{subfig:fit2}]{%
		\includegraphics[width=0.25\textwidth]{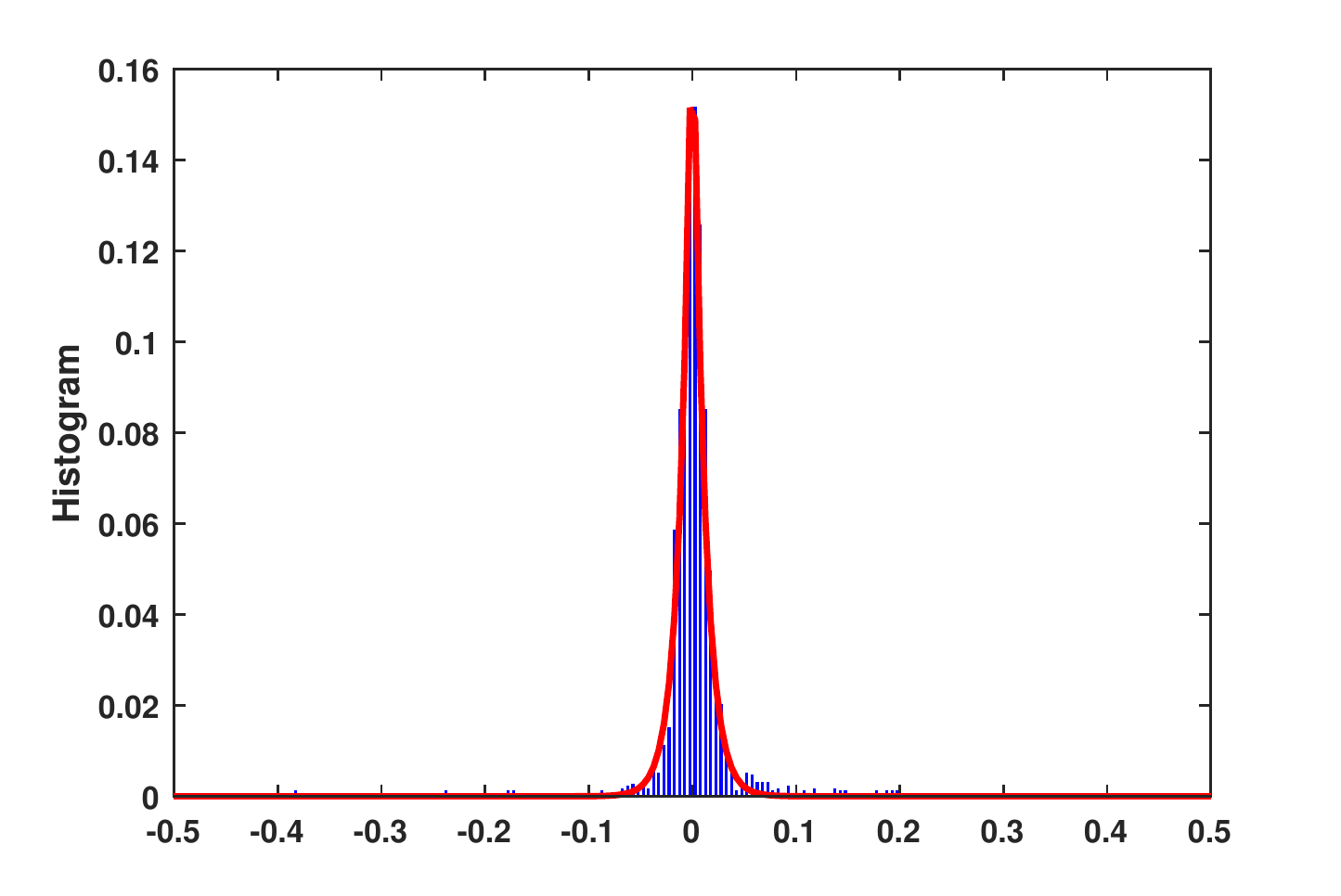}
	}
	\\
	\subfloat[~\label{subfig:fit3}]{%
		\includegraphics[width=0.25\textwidth]{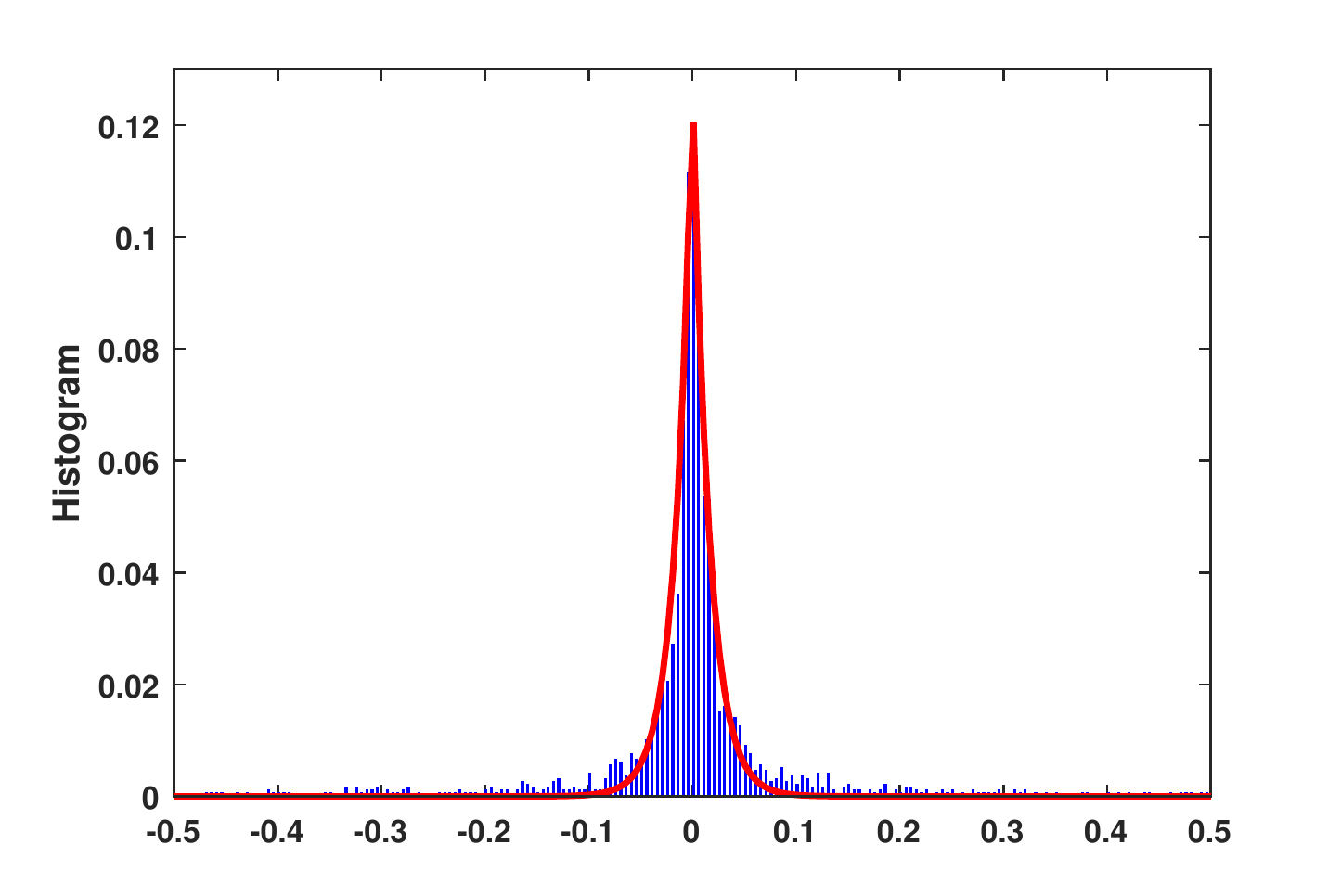}
	}
	\hspace{-0.58 cm}
	\subfloat[~\label{subfig:fit4}]{%
		\includegraphics[width=0.24\textwidth]{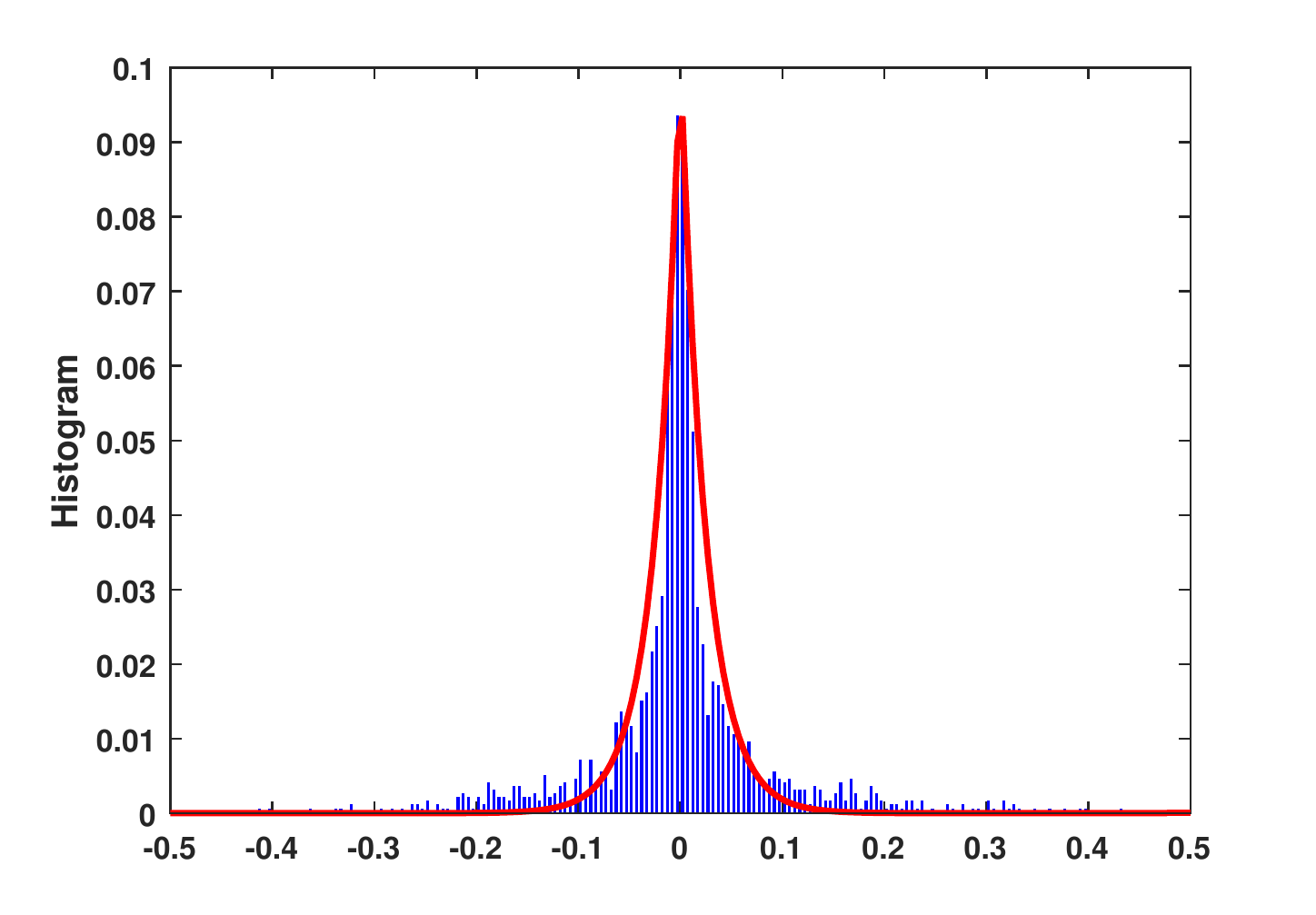}
	}
	\hspace{-0.58 cm}
	\subfloat[~\label{subfig:fit4}]{%
		\includegraphics[width=0.25\textwidth]{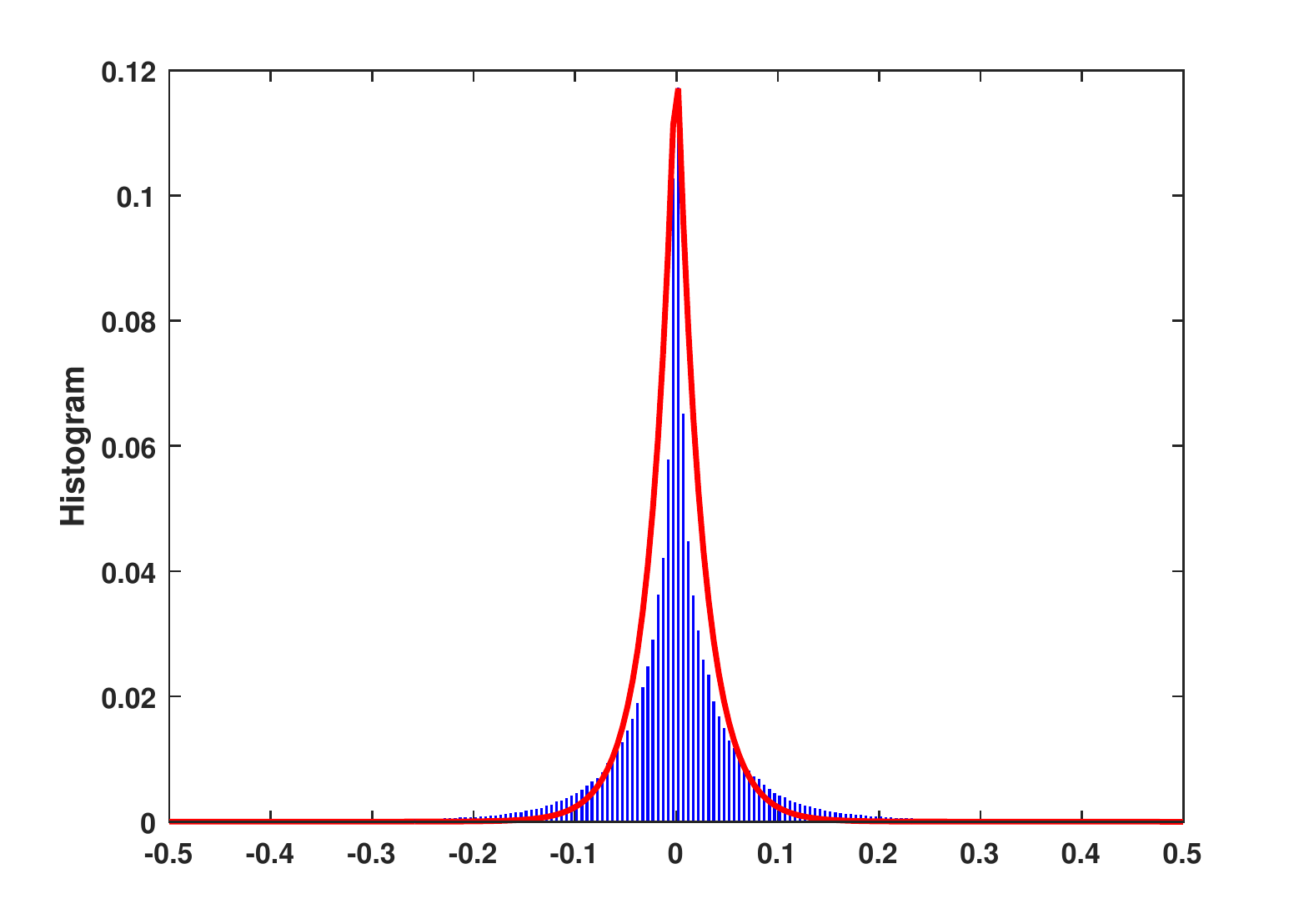}
	}
	\caption{The histogram of the middle frequency DCT coefficients and their approximated with Laplacian distribution for a) {\it{Man}}, b) {\it{Lena}}, c) {\it{Barbara}}, d) {\it{Couple}}, and e) all images in dataset.}\label{fig:lapfit}
\end{figure}

The histogram of the middle frequency DCT coefficients and their corresponding Laplacian approximated distributions are presented in Fig. \ref{fig:lapfit}. According to Fig. \ref{fig:lapfit}, it is confirmed that the Laplacian distribution is a fair approximate distribution for the DCT coefficients.

We will also investigate the drawback of the model in which the DCT coefficients are roughly modeled with Gaussian distribution in Subsection \ref{subsec:noiseattack}. 

\subsection{Transparency of the Proposed Watermarking System}\label{subsec:transparency}
In this subsection, the transparency of the proposed watermarking method is evaluated in terms of Mean Squared Error (MSE), PSNR, and Document to Watermark Ratio (DWR).

Note that the DCT is a unitary transform and obeys Parseval's theorem. So the energy of the noise does not change under this transform. Using this fact, we can analytically calculate the MSE and PSNR for the embedding rule (\ref{eq:embd}) as follows:

\begin{align}
\begin{split}
MSE = &\mathbb{E}\left\{{ (\mathbf{y} - \mathbf{x})^T(\mathbf{y} - \mathbf{x})      }\right\} \\
=& \mathbb{E}\left\{{ [\alpha (2b-1) \mathbf{x}]^T[\alpha (2b-1) \mathbf{x}]      }\right\} \\
=& \alpha ^2 \mathbb{E}\left\{{  \mathbf{x}^T\mathbf{x}     }\right\}=\frac{2N\alpha^2}{\lambda_x^2}.
\end{split}
\end{align}
Hence, for PSNR and DWR, we have
\begin{align}
\begin{split}\label{eq:dwr}
PSNR = & 10\log \left({ \frac{255^2}{MSE}   }\right) = 20\log \left({ \frac{255\lambda_x}{\alpha \sqrt{N}}   }\right),\\
DWR = & 10\log \left({ \frac{\mathbb{E}\left\{{  \mathbf{x}^T\mathbf{x}     }\right\}}{MSE}   }\right) = -20\log \left({ \alpha   }\right).
\end{split}
\end{align}
We can also estimate $\lambda_x$ using the Maximum Likelihood Estimator (MLE) as follows:
\begin{equation}\label{eq:mle}
\hat{\lambda}_x = \mathop {\arg \max }\limits_{ \lambda }{f_{\mathbf{x}}(\mathbf{x}|\lambda)}.
\end{equation}
The solution of (\ref{eq:mle}) is so straightforward. The MLE of $\lambda_x$ can be found as
\begin{equation}
\hat{\lambda}_x = \frac{N}{\sum\limits_{i=1}^{N}{|x_i|}}.
\end{equation}
We use above estimation in our simulations. 

\begin{figure}[b!]
	\centering
	\includegraphics[width=1.05\linewidth]{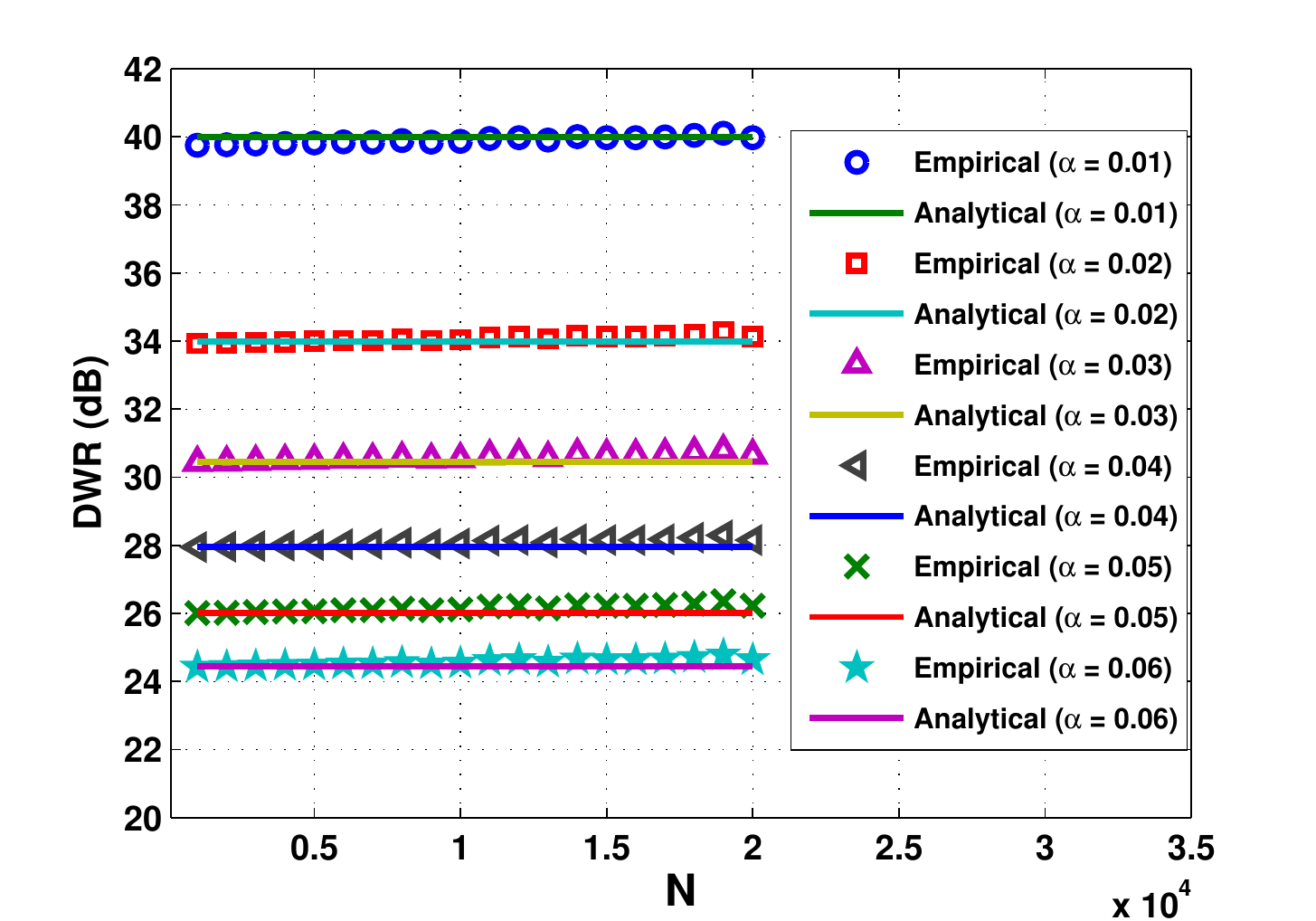}
	\caption{Empirical and analytical values of DWR for different values of $N$ and strength factor.}
	\label{fig:dwr}
\end{figure}

As we expected, in multiplicative spread spectrum embedding rule, the resultant distortion depends on power of the host signal. According to (\ref{eq:dwr}), for a fixed $\alpha$, it is clear that we can increase the transparency of the watermarked signal if we embed the watermark data into the host signal with higher power. Equation (\ref{eq:dwr}) indicates that the DWR does not depend on $N$ in this method. It only depends on strength factor, $\alpha$. We verify this fact by simulation. Several images are watermarked with different values of $N$ and strength factor and the results are shown in Fig. \ref{fig:dwr}. It is obvious that empirical results are very close to the analytical results.

We also watermarked several images with different values of strength factor and the obtained PSNRs are reported in Table \ref{tab:psnr}. As expected, the PSNR decreases as the strength factor increases.
\begin{table}[t!]
	\renewcommand{\arraystretch}{1.20}
	\caption{PSNR (dB) results of the proposed watermarking method with $N=8000$.}\label{tab:psnr}
	\centering
	\begin{tabular}{|c||c||c|}
		\hline
\textbf{Image}	& $\mathbf{\alpha = 0.01}$ & $\mathbf{\alpha = 0.02}$ \\ \hline \hline
\textbf{Girl} 	& 47.94                 & 42.05                     \\ \hline
\textbf{Man}	 & 46.90                 & 41.02                    \\ \hline
\textbf{Couple}	 & 45.88                 & 39.87                    \\ \hline
\textbf{Stream-bridge}  & 45.17       & 39.91                    \\ \hline
\textbf{Lena}     & 45.01                 & 39.20                   \\ \hline
\textbf{Barbara}  & 44.75                 & 38.07                    \\ \hline
\textbf{Elaine}   & 44.11                 & 38.65                    \\ \hline
\textbf{Airplane} & 42.74                 & 36.73                    \\ \hline
\textbf{Aerial}   & 42.22                 & 36.54                      \\ \hline
	\end{tabular}
\end{table}

\begin{figure}[h!]
	\centering
	\subfloat[~\label{subfig:org}]{%
		\includegraphics[width=0.23\textwidth]{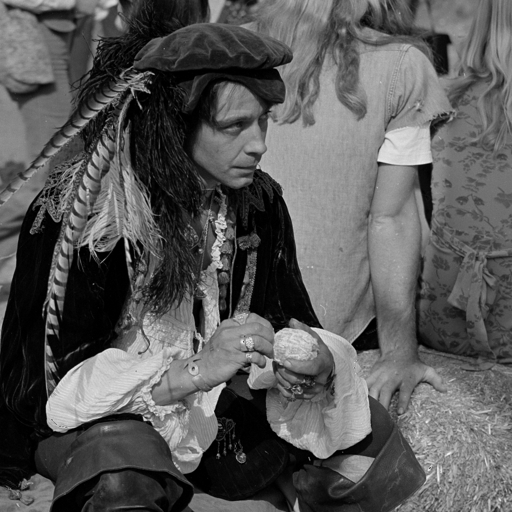}
	}
	\hspace{0.1 cm}
	\subfloat[~\label{subfig:wat}]{%
		\includegraphics[width=0.23\textwidth]{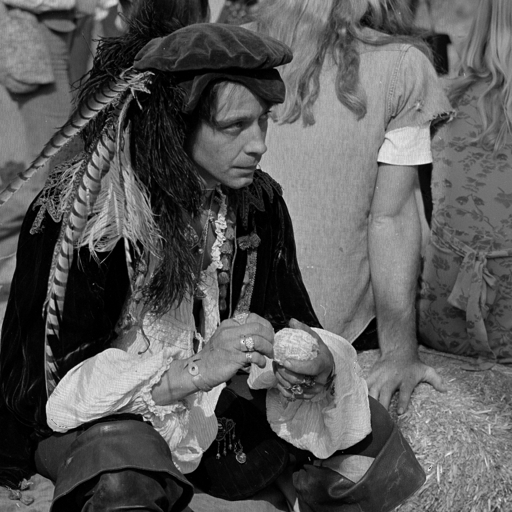}
	}
	\caption{An example of a) original and b) watermarked image. ($N = 8000 $ and $\alpha = 0.02$, and PSNR$=42.03 dB$)}
	\label{fig:orgwat}
\end{figure}

In image and video processing, $37dB$ PSNR is acceptable \cite{ref:kimpsnr}. On average, for $\alpha=0.01$ and $\alpha=0.02$, we get PSNR of $45.04$ and $39.21dB$, respectively, over the image dataset. Moreover, an example of original and watermarked image is shown in Fig. \ref{fig:orgwat}.

According to the results of Table \ref{tab:psnr} and Fig. \ref{fig:orgwat}, we see that our watermarking method has an acceptable level of transparency required for watermarking applications.

\subsection{Noise Attack}\label{subsec:noiseattack}
\begin{figure}[t!]
	\centering
	\includegraphics[width=1.05\linewidth]{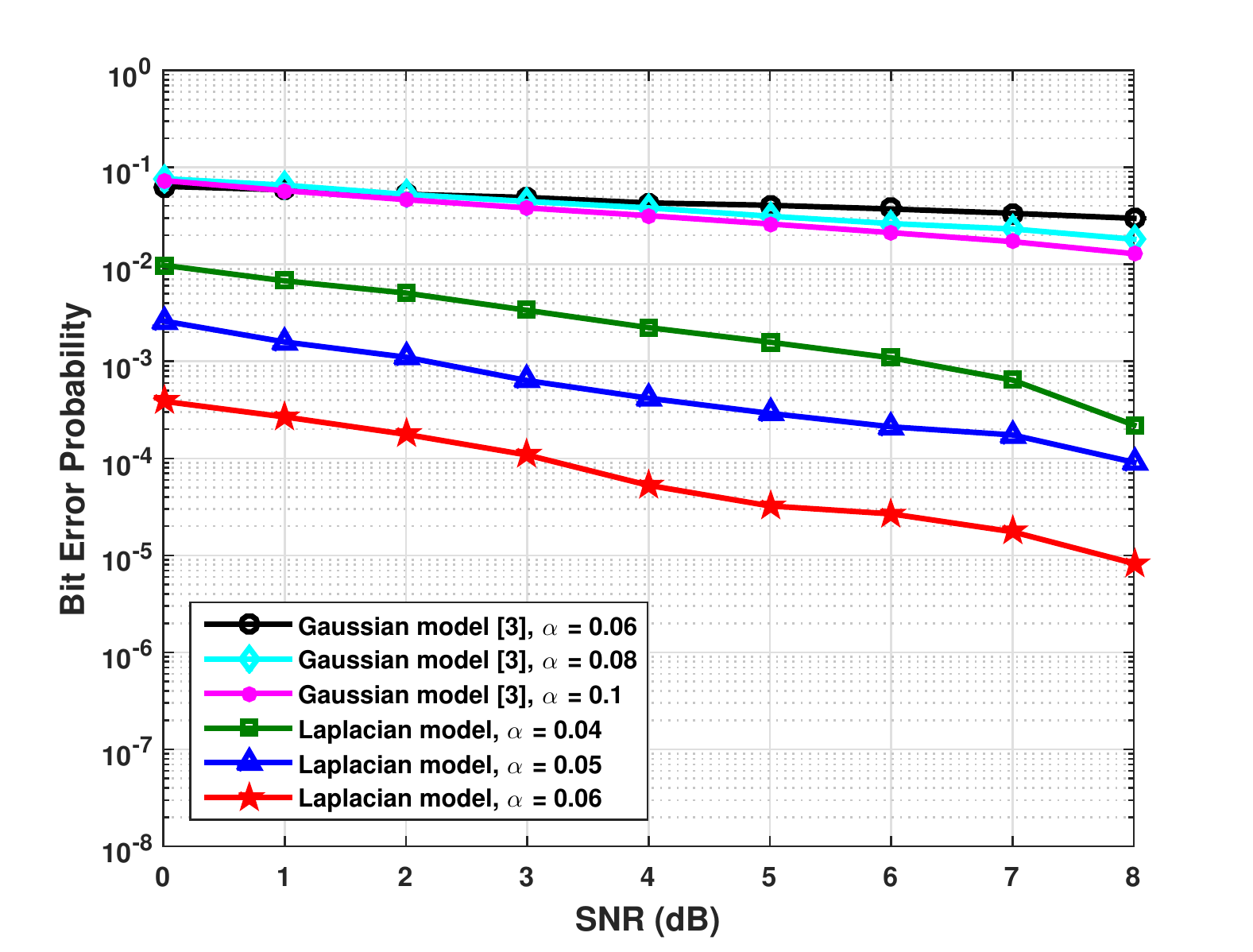}
	\caption{BER of the Laplacian and Gaussian models.}
	\label{fig:treedecodrs}
\end{figure}

In this subsection, we add noise to the watermarked signal and evaluate the performance of the proposed watermarking system in terms of Bit Error Rate (BER) and compare it with the case in which both signal and noise are modeled with Gaussian distributions \cite{ref:cox-spread} for different values of $\alpha$.
Fig. \ref{fig:treedecodrs} shows the BER of the Gaussian and Laplacian model versus Signal to Noise Ratio (SNR). It can be seen that the BER decreases as SNR increases.

Although we used larger strength factors for the Gaussian model, its performance is much worse than the performance of the Laplacian model. For example, the BER of the Laplacian and Gaussian models are, respectively, $5.22\times 10^{-5}$ and $4.4\times 10^{-2}$, for $\alpha = 0.06$ and $SNR=4dB$. 

\begin{figure}[h!]
	\centering
	\includegraphics[width=1.05\linewidth]{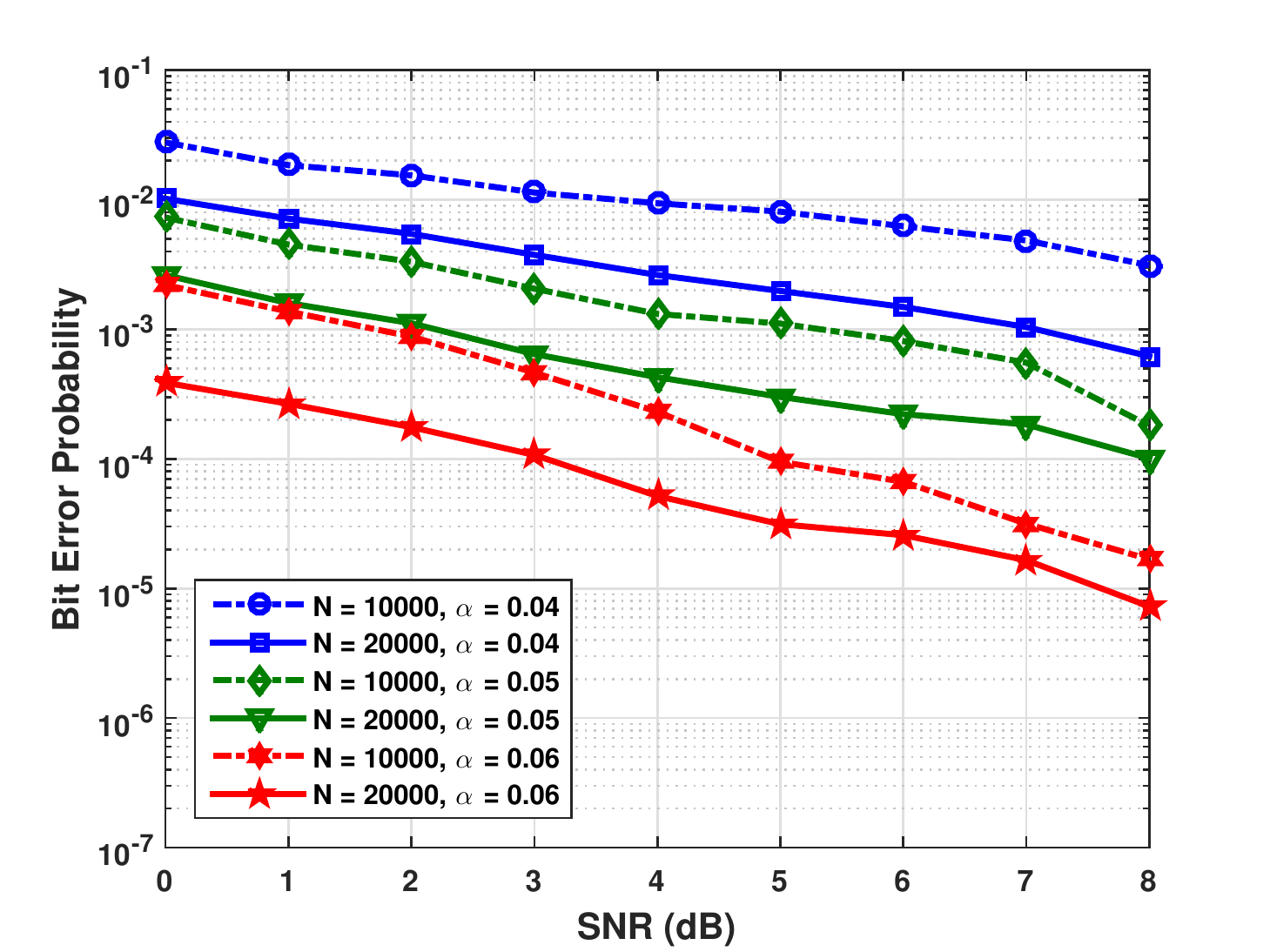}
	\caption{BER of the proposed method with different values of $N$ and $\alpha$.}
	\label{fig:effectn}
\end{figure}

We also investigate the effect of $N$ and $\alpha$ on performance of our method and the results are shown in Fig. \ref{fig:effectn}.

As expected, the BER decreases when $N$ or $\alpha$ increases. However, there is a trade-off between the capacity, transparency, and BER in each watermarking system and the available parameters are set based on the desired application. For example, with $N=10000$ and $\alpha=0.06$ we can get acceptable BER.

\subsection{Pixel Loss Attack}\label{subsec:pixloss}
\begin{figure}[t!]
	\centering
	\subfloat[~\label{subfig:watermarked}]{%
		\includegraphics[width=0.23\textwidth]{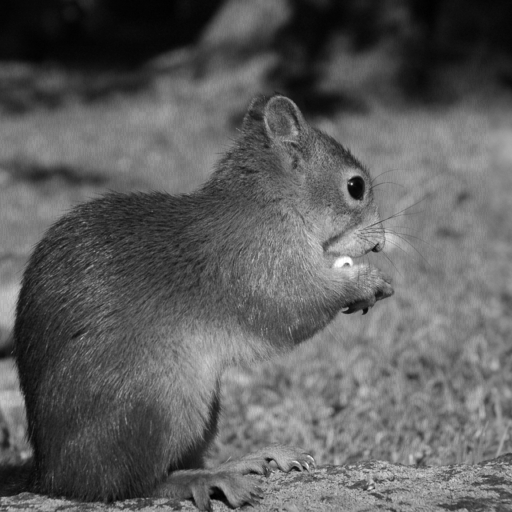}
	}
	\hspace{0.1 cm}
	\subfloat[~\label{subfig:attacked}]{%
		\includegraphics[width=0.23\textwidth]{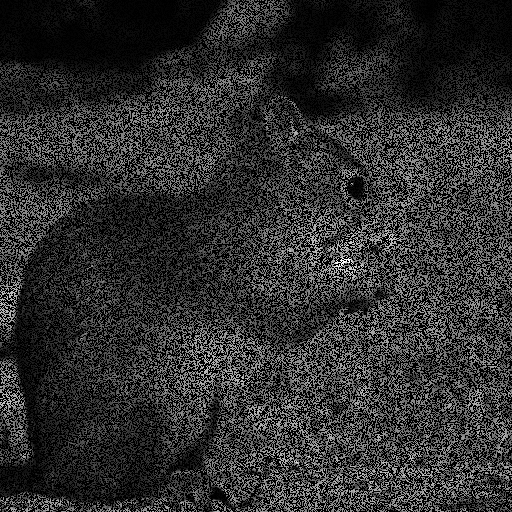}
	}
	\caption{An example of pixel loss attack: a) Watermarked and b) Attacked images. ($P_{loss} = 50\%$ and watermark recovery rate$=97.34\%$)}
	\label{fig:expixloss}
\end{figure}

The performance of the proposed method is also evaluated under pixel loss attack. In this part, the watermark recovery rate is measured after $P_{loss} \%$ of pixels are removed. An example of this attack is shown in Fig. \ref{fig:expixloss}. As seen, while $50 \%$ of the image pixels are removed the proposed method has recovered $97.34 \%$ of the information bits.

\begin{figure}[h!]
	\centering
	\includegraphics[width=1.05\linewidth]{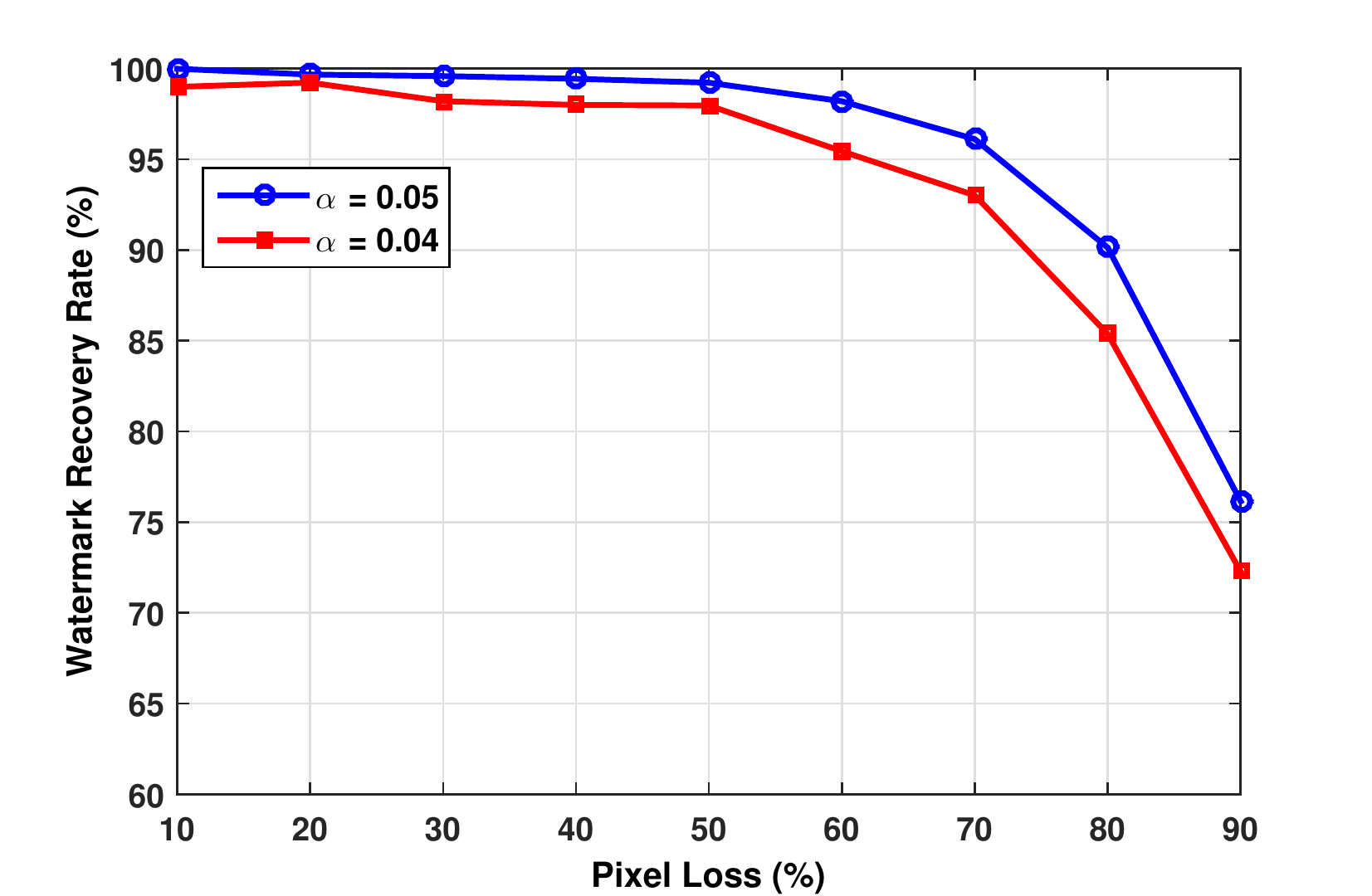}
	\caption{Watermark recovery rate of the proposed method under pixel loss attack.}
	\label{fig:pixloss}
\end{figure}

This attack is also applied on a large dataset and the results are shown in Fig. \ref{fig:pixloss}. In case of pixel loss, we can also use some image processing algorithms for recovery of missing samples (pixels) before extraction of watermark. This may improve the recovery rate.

\subsection{JPG Compression Attack}\label{subsec:jpgcomp}
In this subsection, we present the robustness of the proposed method against the JPG compression attack. For this purpose, we have compressed the embedded image by JPG compression standard with different quality factors and tried to extract the embedded watermarks. We have chosen $\alpha$ in order to get PSNR of $33dB$ between the watermarked and original image. The results are shown in Table \ref{tab:jpg-attack}. You can see that the robustness of the proposed method against JPG compression is acceptable. This is due to that we are choosing the middle frequency DCT coefficients for embedding.

\begin{table}[t!]
	\renewcommand{\arraystretch}{1.20}
	\caption{BER results against JPG compression.}\label{tab:jpg-attack}
	\centering
	\begin{tabular}{|c||c|c|c|}
		\hline
		\textbf{Quality factor}	& 75 & 85 & 95 \\ \hline 
		\textbf{BER ($\%$)}	& 24.3 & 16.5 & 12.1\\ \hline  
	\end{tabular}
\end{table}

\subsection{Brightness Change Attack}\label{subsec:brightness}
We have also evaluated the robustness of the proposed method when the brightness of the watermarked image is changed. Table \ref{tab:brightness} shows the results for different brightness change ratio. 

\begin{table}[t!]
	\renewcommand{\arraystretch}{1.20}
	\caption{BER results against brightness change attack.}\label{tab:brightness}
	\centering
	\begin{tabular}{|c||c|c|c|c|}
		\hline
		\textbf{Brightness change ratio}	& 0.7 & 0.9 & 1.1 & 1.2 \\ \hline 
		\textbf{BER ($\%$)}	& 7.2 & 5.6 & 2.7 & 5.9 \\ \hline 
	\end{tabular}
\end{table}

\subsection{Image Auto Adjustment Attack}\label{subsec:imadjust}
One of the most common image edition is auto adjustment in which the intensity, contrast and brightness of the image will be improved. We have also applied auto adjustment on the watermarked image and then tried to extract the watermark bits. This test has been done on all images in the dataset and the average BER was $2.6\%$.

\section{Conclusion}\label{sec:conclusion}
In this paper, we investigated a multiplicative watermarking method for the image in which both signal and noise were modeled with Laplacian distribution. We derived the optimum decoder for the presented embedding rule. The transparency and performance of the proposed method were evaluated. The simulation results indicated that the proposed method has enough transparency required for watermarking applications. We also compared the optimum decoder with the usual decoder in which both signal and noise are supposed to be distributed with the Gaussian distribution. The BER, in this case, is about $1.1\times10^{3}$ times worse than the true modeling for $\alpha=0.06$ and $SNR=5dB$.


\ifCLASSOPTIONcaptionsoff
  \newpage
\fi
\bibliographystyle{ieeetran}
\bibliography{Citations}

\end{document}